\documentclass[prb,twocolumn,english,superscriptaddress,floatfix,longbibliography]{revtex4-2}
\usepackage[T1]{fontenc}
\usepackage[latin9]{inputenc}
\usepackage{array}
\usepackage{amsmath}
\usepackage{amssymb}
\usepackage{graphicx,hyperref}
\usepackage{wasysym}
\usepackage{txfonts} 
\usepackage{oplotsymbl}
\usepackage{xspace}  

\def\sH{\mathcal{H}}

\def\sR{\mathcal{R}}
\def\sL{\mathcal{L}}
\def\sJ{\mathcal{J}}
\def\sZ{\mathcal{Z}}

\def\sX{\mathcal{X}}
\def\sI{\mathcal{I}}

\def\sU{\mathcal{U}}
\def\sV{\mathcal{V}}

\def\III{{I\!I\!I}}
\def\ny{{\tilde n}}

\def\dR{\mathbbm{R}}

\def\a{\alpha}

\def\e{\epsilon}

\def\ve{\varepsilon}

\def\f{\phi}

\def\vf{\varphi}

\def\kt{\rangle}
\def\br{\langle}
 
\newcommand{\kb}[2]{\ket{#1}\bra{#2}}

\def\sq{\sqrt{\mathrm{TI}}}

\def\Jp{J^\prime}
\def\Jpsq{J^{\prime 2}}

\makeatletter

\pdfpageheight\paperheight
\pdfpagewidth\paperwidth

\usepackage{physics,bbm}
\usepackage{braket}

\def\vf{\varphi}
\def\uno{\mathbbm{1}}

\newcommand{\dnc}{
  {\mathchoice{\displaystyle\raisebox{-1.2pt}{\rhombus}\hspace{-0.4pt}\textnormal{-}}
             {\raisebox{-1.2pt}{\rhombus}\hspace{-0.4pt}\textnormal{-}}
             {{\scriptstyle{\raisebox{-0.22pt}{\tiny\rhombus}\hspace{-0.4pt}\textnormal{-}}}}
             {{\scriptstyle{\raisebox{0pt}{\tiny\rhombus}\hspace{-1.3pt}-}}}}
             }%

\usepackage[export]{adjustbox}
\let\oldincludegraphics\includegraphics
\renewcommand\includegraphics[2][]{%
  \oldincludegraphics[#1,max width=\linewidth,max height=\textheight]{#2}
}

\usepackage{xcolor} 
\usepackage{xfrac} 

\makeatother

\usepackage{babel}
\begin{document}
\title{Hidden topology in flat-band topological insulators: Strong, weak, and square-root topological states}

\author{Juan Zurita}
\email{juzurita@ucm.es}
\affiliation{Instituto de Ciencia de Materiales de Madrid (ICMM), CSIC, Cantoblanco,
E-28049 Madrid, Spain}
\affiliation{Departamento de F\'isica de Materiales, Universidad
Complutense de Madrid, E-28040 Madrid, Spain}
\author{Charles E. Creffield}
\affiliation{Departamento de F\'isica de Materiales, Universidad
Complutense de Madrid, E-28040 Madrid, Spain}
\author{Gloria Platero}
\affiliation{Instituto de Ciencia de Materiales de Madrid (ICMM), CSIC, Cantoblanco,
E-28049 Madrid, Spain}
\begin{abstract}
In this paper, we study a previously unexplored class of topological states protected by hidden chiral symmetries that are local, that is, that protect against any off-diagonal disorder. We derive their related topological invariant for the first time, and show that these previously unidentified symmetries can act together with standard chiral symmetries to increase the protection of the end modes, using the Creutz ladder as an example. Finally, thanks to local hidden symmetries, we show that the diamond necklace chain can have three different types of topological end modes: strong, weak or square-root, with some of the states inheriting their topology from others. We therefore show that a square-root topological insulator can be identified as its own parent.
\end{abstract}
\maketitle

\section{Introduction}
Topological insulators (TIs) have been extensively studied in recent years, due to the robustness of their boundary modes against disorder, while still being fairly accessible models to implement and control. Due to these properties, they are important assets in the design of robust quantum information protocols like quantum state transfer \cite{Stace2004,Estarellas2017,Lang2017,Zurita2023} and entanglement distribution \cite{Benito2016,Yang2016a,Hu2020,ZuritaEnt}. Moreover, it has been shown that a huge percentage of all possible atomic structures are topological, many of which were not previously identified \cite{Bradlyn2017}. This proved that the traditional methods of identification and classification of topological materials are clearly insufficient.

In the study of topological models at the single-particle level, the first step taken is often the reduction of the Hamiltonian to its bulk reciprocal version, an $N_c\times N_c$ matrix defined by the partial inner product $\sH(k)=\bra{k}\sH\ket{k}$, where $N_c$ is the number of atoms in a minimal unit cell \cite{Asboth2015}. Only then are the protecting symmetries explored on each $k$ subspace. This approach, while usually satisfactory, implicitly assumes the protecting symmetries commute with all lattice translations, that is, that they have the same periodicity as the lattice. Extensive research has been made lately in a variety of cases where other lattice symmetries that mix several $k$ subspaces can protect the end modes, grouped under the name of \textit{crystalline topology}. These symmetries, although interesting in their own right, are easily broken by local disorder \cite{Fu2011,Hsieh2012}.

The symmetries that cannot be found directly in $\sH(k)$ are often referred to as \textit{hidden} \footnote{The meaning of the term varies from work to work, for example, some authors \cite{Hou2013,Hou2016} reserve the name for antiunitary operators.} \cite{Hou2013,Cariglia2014,Fukui2013,Li2015B,Hou2016,Zurita2021,Zuo2024}. In this paper, we focus on local hidden (LH) symmetries, which were first alluded to in Ref. \cite{Zurita2021} and are capable of true protection against local off-diagonal disorder. Their operators are diagonal, and they are hidden because their periodicity is different from that of the lattice, not because they mix states with different $k$ by rotating or reflecting the crystal, like in crystalline TIs.

We explore the properties of two models with LH symmetries: the Creutz ladder (CL) and the diamond necklace chain (DNC). We derive and calculate their hidden topological invariants for the first time, and show that LH and standard symmetries can work together to increase the protection of the end modes. We also explore the connection between \textit{hidden topology} and other types of exotic topology, like square-root or weak end modes. Both of these states inherit some degree of topological protection from the states in other models, dubbed their \textit{parents} \cite{Padavic2018,Kremer2020,Marques2021}. As we show, hidden symmetries allow the DNC to be identified as its own parent, with its outer gap states inheriting the topology from the ones in its central gap. We thus demonstrate that identifying local hidden symmetries correctly is crucial in order to properly describe some exotic topological models.

\begin{figure}[!tph]
\mbox{%
\includegraphics[width=1\columnwidth]{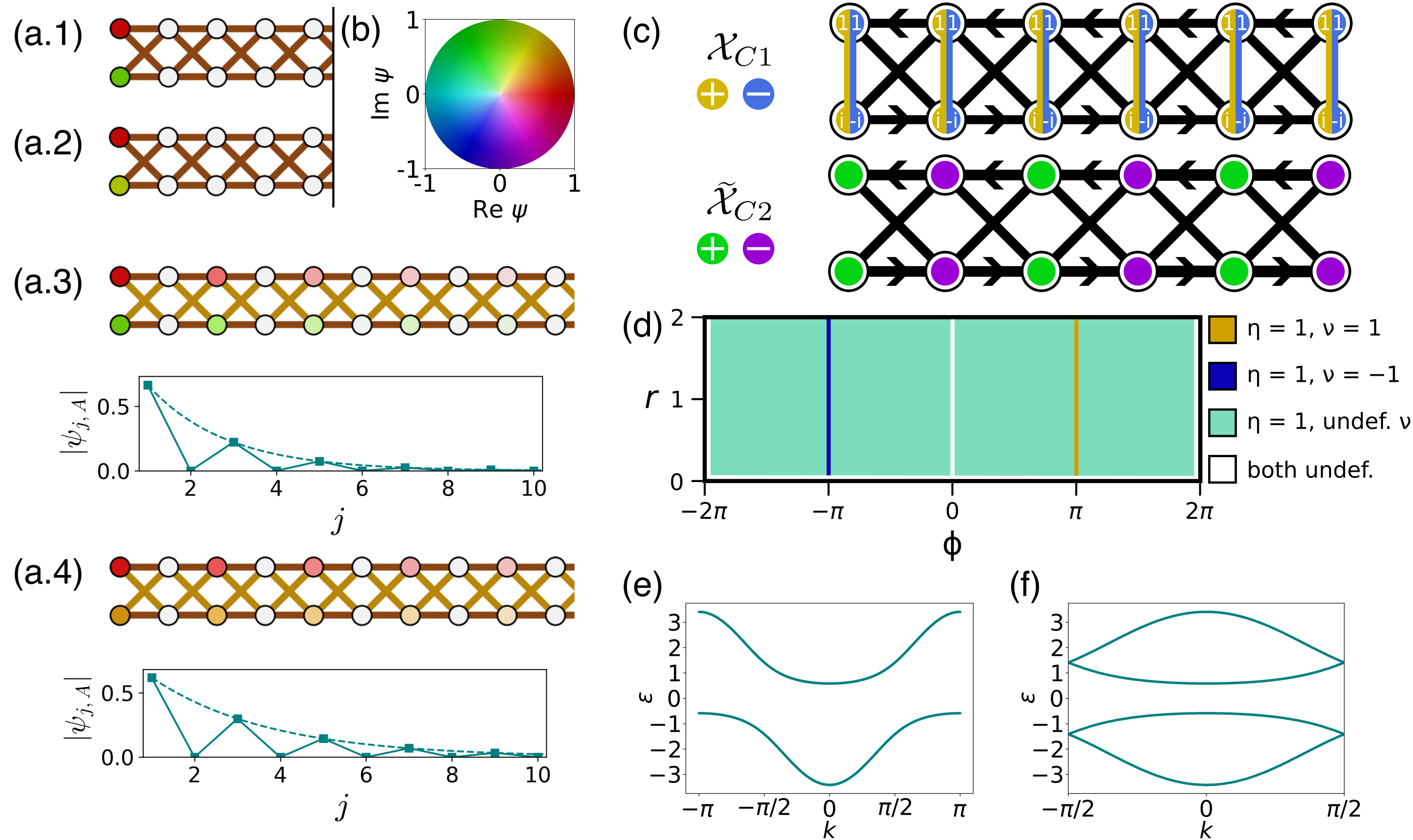}%
}\caption{(a) Hidden-chiral left end modes for different values of $(\phi,r)$: (a.1) $(\pi,1)$, (a.2) $(5\pi/4,1)$, (a.3) $(\pi,0.5)$, and (a.4) $(5\pi/4,0.5)$. All pictures show the wavefunction following the legend in (b). The two latter cases include the absolute value of the wavefunction on the $A$ sites (equal to that in $B$ sites), as well as its exponential envelope. (c) Eigenstates (chiralities) of the two chiral operators mentioned in the main text. (d) Creutz ladder phase diagram as a function of $(\phi,r)$. The two $\sX_{C1}$-chiral lines at $\phi=\pm\pi$ are included inside the larger $\sX_{C2}$-chiral region, in green. We show the value of the winding number $\nu$ and the hidden winding number $\eta$. (e,f) Bands of the $(\pi/2,1)$ CL. The chiral symmetry is not apparent in the Brillouin zone using a two-site unit cell (e), but it appears when taking the four-site unit cell (f), plotted here as a function of the two-site-cell momentum $k$. \label{fig:CL}}
\end{figure}

\section{Hidden topology: the Creutz ladder}
\subsection{Hidden symmetries}
If a Hamiltonian $\sH$ commutes with a unitary operator $U$, a basis of eigenstates of both operators exists, $B_{\sH,U}$. In particular, if $\sH$ is a 1D Hamiltonian with periodic boundary conditions and $U=T_1$ is a spatial translation of one unit cell in the positive direction, each of the states in $B_{\sH,U}$ will have a well-defined energy $\ve$ and quasimomentum $k$, the eigenvalues of $\sH$ and $T_1$. This is the basis for Bloch's Theorem.

Therefore, the matrix $\sH$ can be block-diagonalized into blocks of well-defined $k$, so the subspace spanned by each block is decoupled from the rest. These blocks are the reciprocal bulk Hamiltonians $\sH(k)$, for each allowed $k$.

This is the justification often used to limit the study of topological Hamiltonians to $\sH(k)$. While this is true from the point of view of particle dynamics, a topological study might require a more global perspective.

To illustrate why this is the case, we note that a Hamiltonian with two different symmetries $U_1$ and $U_2$ can only be block-diagonalized into blocks with defined eigenvalues $u_1$ and $u_2$ if the two symmetry operator commute with each other.

A slightly different situation arises when an operator $\sX$ \textit{anticommutes} with $\sH$ (as required for a chiral symmetry). Then, $\sH$ can always be \textit{block-off-diagonalized} into two off-diagonal blocks of equal size, leaving the two blocks in the diagonal empty. If a Hamiltonian commutes with $T_1$ and anticommutes with $\sH$, and $[\sX,T_1]=0$, then each of the blocks $\sH(k)$ induced by $T_1$ can be off-block-diagonalized using $\sX$. This operation is required to obtain a winding number from $\sH(k)$.

If $\sX$ does not commute with $T_1$ (i.e. is not periodic in the lattice), the topology of the Hamiltonian cannot be found directly in momentum space. Different approaches have been used in the literature to study these situations \cite{Hou2013,Cariglia2014,Fukui2013,Li2015B,Hou2016,Zurita2021,Zuo2024}, but in this work we aim for a complete understanding of these phases by finding associated topological invariants. 

\subsection{The Creutz ladder}
We consider the rungless Creutz ladder, described by the Hamiltonian \cite{Creutz1999}:
\begin{align}
    \sH_{CL} = \sum_{j=1}^{N-1}\sum_{\a=A,B}&\left[-Je^{is_a\phi /2}    \kb{j+1,\a}{j,\a}  \right.\nonumber\\
    &\left.\vphantom{e^{is_a\phi /2}}-rJ\kb{j+1,\overline\a}{j,\a} + H.c.\right],
\end{align}
where $j=1,\ldots,N$ labels the unit cells, $\a=A,B$ labels the two atoms in each cell, $\overline A = B$ and vice versa, and $s_{A/B}=\pm 1$. The horizontal and diagonal hopping amplitudes are $J$ and $rJ$, while $\f$ is the magnetic flux through any planar four-link loop.

This model is a topological insulator in class BDI, with end modes protected by the (non-local) chiral symmetry $\sX_{C1}=\sigma_y$, if $\phi = \pm \pi$, with a winding number of $\nu = \pm 1$. The end modes are chiral with respect to this operator, that is, each of them only has support on one of the $|A\kt \pm i |B\kt$ subspaces. Given that the operator is not diagonal (not local), it does not protect against off-diagonal disorder.

That is the usually studied regime. However, it was recently shown \cite{Zurita2021} that the case with general $\f$ is also topologically protected, as long as $\f \ne 0 \mod 2\pi$ (in which case the bands touch). In this case, the hidden chiral symmetry protecting the end modes is the $2N\times 2N$ matrix:
\begin{equation}
    \sX_{C2} = \textrm{diag}_{2N}(1,1,-1,-1,\ldots), \label{eq:XC2}
\end{equation}
which repeats every four rows and acts on the full space.

The eigenstates of this operator are the single-site states $|j,\a\kt$, with their chirality being $(-1)^{j+1}$. This makes the symmetry local, and means that chiral end modes in this regime would only have support on either odd or even rungs. As can be seen in Fig. \ref{fig:CL}, the end modes of the system are either compact (for $r=\pm 1$), i.e. localized only on the first and last rung, or have support only on odd or even rungs (if $|r|\ne 1$).

\begin{figure*}[!tph]
\mbox{%
\includegraphics[width=0.8\textwidth]{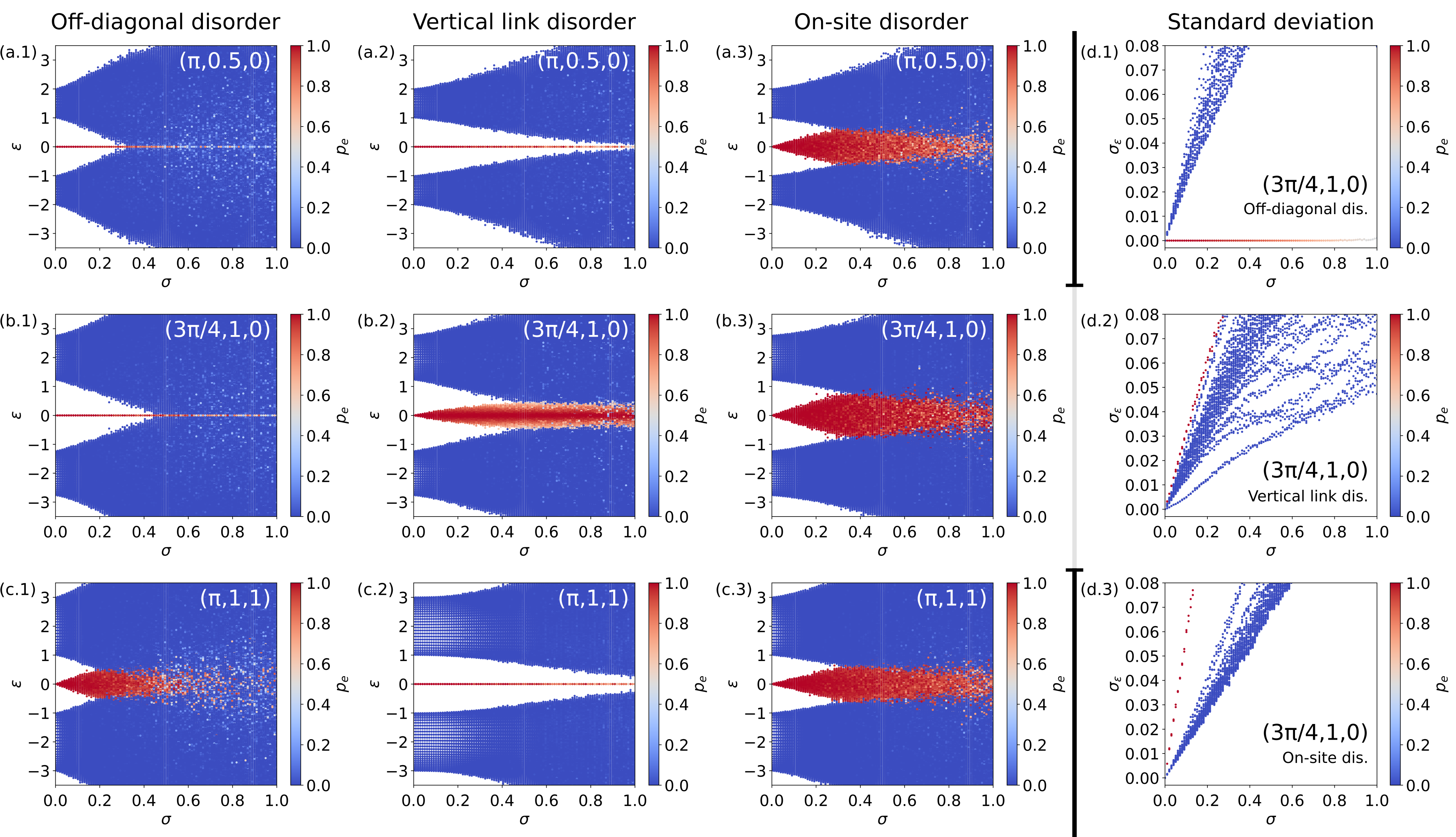}%
}\caption{Disordered Creutz ladder spectra for a finite system with $N=30$ rungs, labeled with $(\phi,r,\epsilon)$. The energies of 1000 realizations were plotted at each disorder strength $\sigma$, with their color representing their probability of finding them in a pristine edge state. We consider off-diagonal (1), vertical link (2) and on-site (3) disorders for the parameter values $\phi=\pi,r=0.5$ (a), $\phi=3\pi/4,r=1$ (b) and $\phi=\pi,r=1$ with an added energy imbalance of $\epsilon=1$ (c). We also include the standard deviations of each energy level in the (b.1-3) cases in (d.1-3). \label{fig:DisCL}}
\end{figure*}

In order to find a suitable topological invariant for this phase, we start from the usual winding number formula \cite{Stanescu2016}:
\begin{equation}
    \nu = \frac{i}{2\pi} \oint_{BZ} \textrm{tr}[q(k)^{-1}\partial_k q(k)] dk, \label{eq:nu}
\end{equation}
where $q(k)$ is the off-diagonal block of the reduced Hamiltonian in the chiral basis,
\begin{equation}
    Q(k) = \left(\begin{matrix}
        0 & q(k) \\
        q(k)^\dagger & 0
    \end{matrix} \right)=\uno - 2P(k), \label{eq:Qk}
\end{equation}
and $P(k)$ is the projector onto the space under the studied gap. In this case, it will preserve the eigenstates of $\sH(k)$ with $\ve<0$.

The reduced Hamiltonian $Q(k)$ can be obtained from $\sH(k)$ by flattening all positive (negative) bands and taking them to an energy of $1$ ($-1$). This is because the specific energy of each $k$-state is irrelevant for the winding number; only the change in the eigenstates when varying $k$ matters. Given that a chiral Hamiltonian is block-off-diagonal in the chiral basis by definition, $Q(k)$ always has the form in Eq. (\ref{eq:Qk}).

The hidden chiral symmetry in Eq. (\ref{eq:XC2}) is periodic every four sites, so we consider a four-site unit cell, $ \{ |j,A\kt,|j,B\kt,|j+1,A\kt,|j+1,B\kt\}_{\textrm{odd }j }$. In this basis, the reciprocal Hamiltonian of the Creutz ladder has the form:
\begin{equation}
    \tilde\sH(\tilde k)=\left( \begin{matrix}
        0 & \tilde h(\tilde k)\\
        \tilde h(\tilde k)^\dagger & 0
    \end{matrix}\right),
\end{equation}
\begin{equation}
     \tilde h(\tilde k)= -2J e^{-i\tilde k/2} \left( \begin{matrix}
        \cos \frac{\f - \tilde k}{2} & r\cos \frac{\tilde k}{2}\\
        r\cos \frac{\tilde k}{2} & \cos \frac{\f + \tilde k}{2}
    \end{matrix}\right),
\end{equation}
where $\tilde k = 2k$ is the four-site-cell momentum. We can see that the Hamiltonian is block-off-diagonal, because we are already using the chiral basis. As is the case in superlattices, the bands fold into a reduced Brillouin zone. There, a new chiral symmetry that was absent from the original two-site Brillouin zone appears, as shown in Fig. \ref{fig:CL} (e,f).

The reduced Hamiltonian $\tilde Q(\tilde k)$ is also block-off-diagonal, with $\tilde q(\tilde k)$ being its upper-right block. We can then calculate the \textit{hidden winding number} $\eta$, by substituting $k$ and $q(k)$ with $\tilde k$ and $\tilde q (\tilde k)$ in Eq. (\ref{eq:nu}). It is the topological invariant associated with $\tilde \sX_{C2}=\textrm{diag}_4(1,1,-1,-1)$.

The integral was numerically evaluated from $-\pi$ to $\pi$, using finite differences and taking care to use the same matrix values for both endpoints. The integrand converges to the constant value of $-i$, making $\eta = 1$ for all values of $(\phi,r)$ as long as $\phi \ne 0 \mod 2\pi$ and $r\ne 0$.

This is a crucial point: what were identified as two topological phases by the standard winding number $\nu$ at the $\sX_{C1}$-chiral lines $(\phi=\pm\pi,r)$, now have the same hidden winding number $\eta$. This makes sense, because the hidden chirality of the end modes (i.e., them being on odd or even rungs) does not depend on the sign of $\f$, as was the case with $\sX_{C1}$. We show the CL topological phase diagram in Fig. \ref{fig:CL} (d).

Finally, in order to illustrate the strong protection of the hidden-chiral states, and compare it to that of the standard chiral states, we calculated the spectrum of $1000$ different disordered CL Hamiltonians at three points in the phase diagram and for different kinds of disorder: off-diagonal (OD), vertical link (VL), and on-site (OS) disorder. Details on its implementation are included in Appendix \ref{sec:A}. OS disorder, as usual, always breaks the protection. The two chiral symmetries $\sX_{C1,2}$ protect against OD and VL disorder, respectively. For $\phi=\pi$, both chiral symmetries are present, and the energy of the end modes is robust against both types of disorder, as shown in Fig. \ref{fig:DisCL} (a). When breaking only $\sX_{C1}$ [\ref{fig:DisCL} (b)] or $\tilde\sX_{C2}$ [\ref{fig:DisCL} (c)], we can see that the modes are protected only against one of the two disorder types. To break $\tilde\sX_{C2}$, we introduce an on-site energy imbalance of $\epsilon=1$ between the two CL legs, which corresponds to on-site energies of $\pm 1$ in the $A/B$ sites. We include the standard deviation of the energies for the (b) case in Fig. \ref{fig:DisCL} (d) (grouping them by their order in the disordered spectrum), in order to show the lack of fluctuations in the protected case.

We want to draw attention to an important point: as shown in Fig. \ref{fig:DisCL} (a), the doubly chiral regions, $(\phi=\pm\pi,r)$, are protected against some type of disorder as long as either of the two symmetries is preserved ($\sX_{C1}$ or $\tilde\sX_{C2}$). This greatly enhances its robustness, and contrasts with the conventional view in the literature of topological insulators, that argues that the presence of two chiral symmetries in a model does not affect its topological properties \cite{Stanescu2016}. Furthermore, the presence of $\tilde\sX_{C2}$ explains the protection against OD disorder of the $\pi$-flux Creutz ladder end modes, which cannot be explained using standard symmetries.

We explore below the properties of a more exotic model, in which hidden symmetries are interweaved with another type of exotic topology: square-root topology. To that end, we now briefly review the main ideas behind the latter.

\section{Weak and square-root topology}
Our next goal is to explore the interplay of hidden symmetries with other, more exotic varieties of topology. However, in order to do this, we first need to familiarize ourselves with them. Therefore, in this section, we present the concepts of \textit{weak} and \textit{square-root topology}.

\subsection{Weak topology: the SSH ladder}
In a finite 1D topological insulator, the end modes at each side of the model will have a finite overlap and hybridize, acquiring a small energy below and below zero and thus creating bonding and antibonding states. These can be used to implement transfer protocols between the two edges.

If the overlap is small, like in a long enough SSH chain or Creutz ladder, the states still remain close to zero energy, thus retaining a moderate amount of topological protection. In a transfer protocol, for example, the overlap must be enough for the process to be fast, but not so large so as to lose all topological protection.

However, if the overlap between the states is large, most of the topological protection is lost. This is so because, strictly speaking, strong topological protection can only pin a state at exactly zero energy when it is isolated (like in a semi-infinite model), given that chiral symmetry does not restrict the hybridization energies between end modes.

A class of topological models exist, weak topological insulators, where the overlap between end modes is unavoidably large. They are usually built by stacking several strong topological systems into a higher dimension, each of them providing an end mode at each boundary. These states will have large overlaps, given that they occupy neighboring sites and have the same analytical expression, and so they hybridize and acquire nonzero energies. These energies are highly dependent on tunneling amplitudes and on-site energies, and so are not protected against disorder in general.

The paradigmatic quasi-1D example of a weak topological insulator (WTI) is the SSH ladder, formed by two SSH chains connected with vertical hoppings \cite{Padavic2018}. We show this model in Fig. \ref{fig:SqBands} (c), as well as its weak edge states. We also include the two SSH chains (strong TIs) that get coupled with the vertical links to produce the ladder. These are the \textit{parent} models of the SSH ladder.

\subsection{Square-root topology: the rhombus chain}
\begin{figure}[!tph]
\mbox{%
\includegraphics[width=0.97\columnwidth]{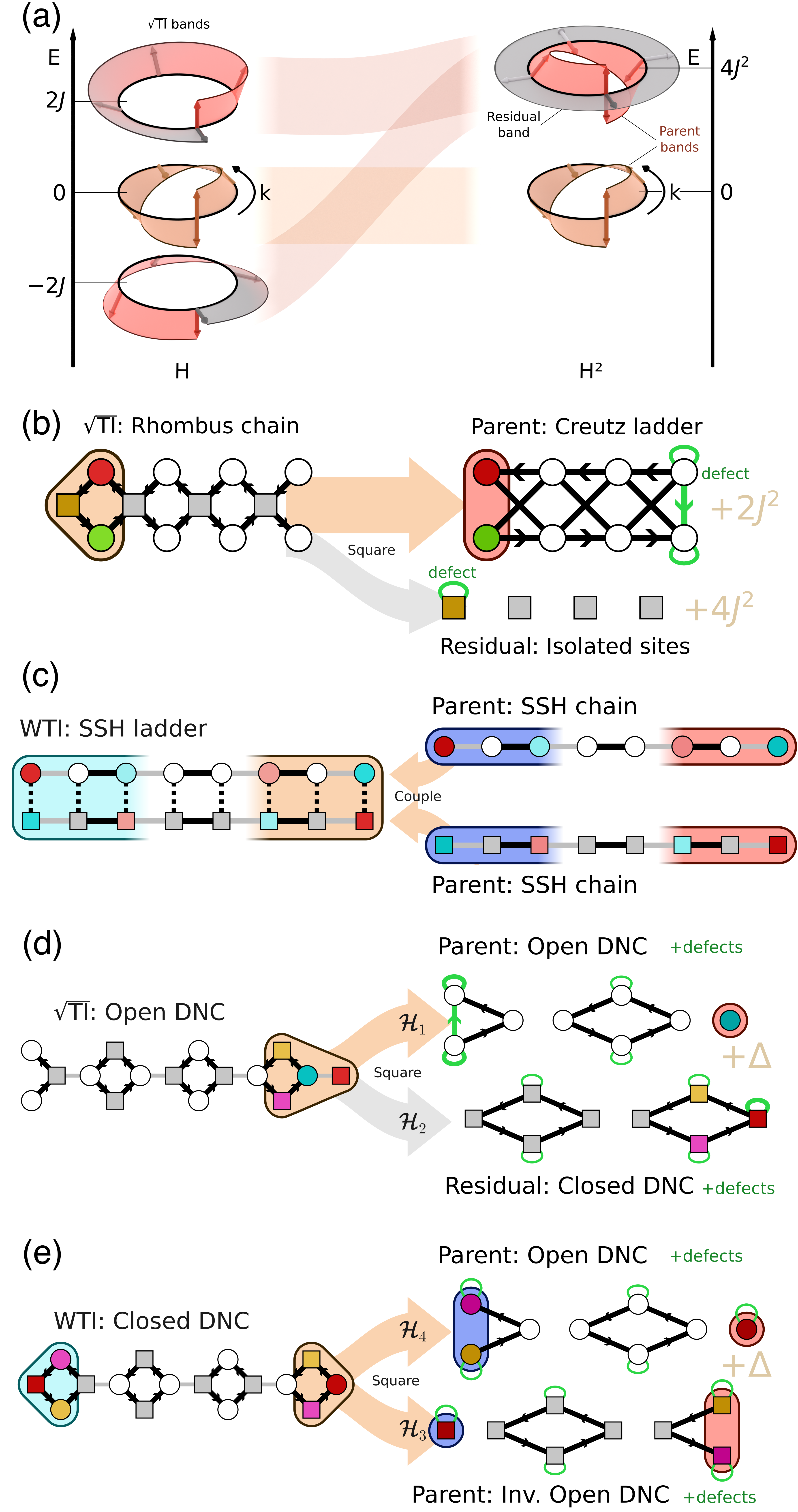}%
}\caption{(a) Effect of the squaring operation on the bands of a square-root topological insulator. We show the topology of each band $n$ by representing the value of the partial Zak phase integral in Eq. (\ref{eq:partialZak}) for each value of $k$. The $\sq$ Hamiltonian (left) has a central topological band at zero energy with a Zak phase of $\pi$, and two bands at $\pm \epsilon$, both with a Zak phase of $-\pi/2$. After squaring (right), the central band does not change, but the outer bands end up degenerate at $\epsilon^2$. Due to the geometry of the lattice and/or magnetic interference, two invariant subspaces are created, shown in grey (residual space, trivial) and red (topological). The original outer bands are colored as a function of the band each state $|u_n(k)\rangle$ ends up in after squaring. (b) Rhombus chain and the two subspaces its square divides into. Defects in the lattices are shown in green, and the global energy shift is indicated for each one. (c) SSH ladder and the two SSH chains that form it. (d) Open DNC and the two subspaces its square divides into, both with a $\Delta$ energy shift. (e) Closed DNC and the two subspaces its square divides into, both with a $\Delta$ energy shift. We show the relevant end modes of each model using the legend from Fig. \ref{fig:CL} (b). We do not plot amplitudes smaller than $0.02$. The full form of the obtained models is included in Appendix \ref{sec:C}.
\label{fig:SqBands}}
\end{figure}

As pointed out recently \cite{Kremer2020,Marques2021}, the $\pi$-flux rhombus chain (RhC), shown on the left in Fig. \ref{fig:SqBands} (b), has left-side edge states in its two gaps that inherit topological properties from a parent topological Hamiltonian. It can be obtained by squaring the RhC Hamiltonian, $\sH_R$, and observing that the resulting Hamiltonian describes two decoupled systems. One of them is a Creutz ladder with some defects on its right side (the parent topological model), and the other is a one-dimensional (1D) chain (the so-called residual model). The defects on the right end of the Creutz ladder, which consist of a different on-site energy and vertical hopping amplitude compared with the other sites, make its right end mode sink into its bands \cite{Marques2021}. This model is shown on the right in Fig. \ref{fig:SqBands} (b). As we show below, this causes the square-root topological ($\sq$) model to have two semi-protected end modes on its left side. These defects stem from the fact that the square-root model is not symmetric under spatial reflection. Its reflected counterpart would, naturally, have the end modes on its right side instead. This fact can always be used to relate these states to a two-dimensional (2D) topological invariant, as we show in Appendix \ref{sec:B}.

The RhC inherits topological features from the CL due to the fact that every eigenstate of $\sH_R$ with energy $\ve$ is also an eigenstate of $\sH_R^2$ with energy $\ve^2$. This has two consequences:
\begin{enumerate}
    \item The bulk bands of the RhC have Zak phases of exactly $(-\pi/2,\pi,-\pi/2)$. The zero energy band stays invariant under the squaring operation and becomes the lower CL band, while the other two bands, originally at energies $\pm\ve$, combine at energy $\ve^2$ to create the upper CL band and the decoupled sites. Considering the geometry of the Bloch bundle formed by the bands at $\ve^2$, we can see how the RhC outer bands, with $\sZ=-\pi/2$, can \textbf{combine to produce a topological band and a trivial band}, with zero and $\pi$ Zak phases, respectively. We illustrate this schematically using the partial Zak phases
    \begin{equation}
        z_n(k) = i\int_0^k \langle u_n(k^\prime)|\partial_{k^\prime} |u_n(k^\prime)\rangle dk^\prime \label{eq:partialZak}
    \end{equation}    
    in Fig. \ref{fig:SqBands} (a), where the twist angle of the arrow at each $k$ in band $n$ is equal to $z_n(k)$. If its arrow stays the same after a roundtrip, the band has $\sZ=0$. If it points in the opposite direction, it has $\sZ=\pi$.
    \item Given that every $\sH^2$ eigenstate at $\ve^2$ is a combination of a subset of the $\sH$ eigenstates at $\pm\ve$, each state of $\sH^2$ is a combination of a few states \footnote{Here, ``a few'' is the total number of states at each pair of energy values, $\pm\ve$. In the flat band case, where the degeneracy is high, we can always choose compact bases, so our claim still holds for model with reasonable-ranged interactions.} in $\sH$ and vice versa. In this light, we can see that \textbf{the left edge state of the parent CL couples to the first decoupled site to produce the two left square-root topological end modes}. No such state exists on the right side due to the defects, so the square-root end modes do not exist there.
\end{enumerate}

These states are reminiscent of those in weak topological insulators, in which several protected end modes combine to produce less protected states \cite{Ringel2012,Padavic2018}, with the difference that one of the states in the mix (coming from the residual model) is not necessarily symmetry-protected.

\section{Square-root and weak hidden topology: the diamond necklace chain}
The diamond necklace chain (DNC) is an extremely interesting model, the topology of which is not straightforward to understand. The DNC itself \cite{Nicolau2023,Kempkes2023a,Burgher2024} and some dimerized variants \cite{Li2021,Kempkes2023a} have been recently studied, with a special focus on its peculiar symmetries. Here, we provide a full analysis of its different types of possible end modes, all of which are topological due its LH chiral symmetry, which was previously unidentified. Its bulk Hamiltonian is:
\begin{align}
    \sH_{\dnc} = \sum_{j=1}^N &\left[ -Je^{i\phi/4}\left(\kb{j,B}{j,A} + \kb{j,D}{j,B} + \kb{j,C}{j,D} +\right.\right.\nonumber\\
    &\left.\left.+ \kb{j,A}{j,C} \right) -\Jp \kb{j+1,A}{j,D}  \right], \label{eq:DNC}
\end{align}
where $J$ is the diamond hopping amplitude, $\Jp$ is the horizontal one, and $\phi$ is the flux through each diamond. We set $J=1$.

The model has four bands, with the LH chiral symmetry:
\begin{equation}
    \tilde\sX_\dnc = \textrm{diag}_8(1,-1,-1,1,-1,1,1,-1),
\end{equation}
which is periodic every two unit cells [see Fig. \ref{fig:DNC} (a)]. This symmetry is diagonal, and so it is not broken under off-diagonal disorder in the links of the lattice.

The end modes that appear in the model depend strongly on the choice of open boundary conditions. The model is always hidden-topological, but in one of the cases there are two strong and two square-root topological states, while, in the other, four weak end modes appear. We first consider the former case.

\subsection{Open DNC}
We first choose the boundary conditions so that the leftmost diamond is open, its right side ending in a $\Jp$ link, as shown on the left in Fig. \ref{fig:SqBands} (d). We call this choice the open DNC. Naturally, leaving the rightmost diamond open leads to the same states, but on the opposite ends.

The model has a left and a right edge state in the central gap, and one additional right end mode in each outer gap. After closer inspection, it can be seen that the states in the central gap are protected against off-diagonal disorder by $\tilde\sX_\dnc$, for any choice of $\Jp$ with a non-zero $\phi$.

\begin{figure}[!tph]
\mbox{%
\includegraphics[width=1\columnwidth]{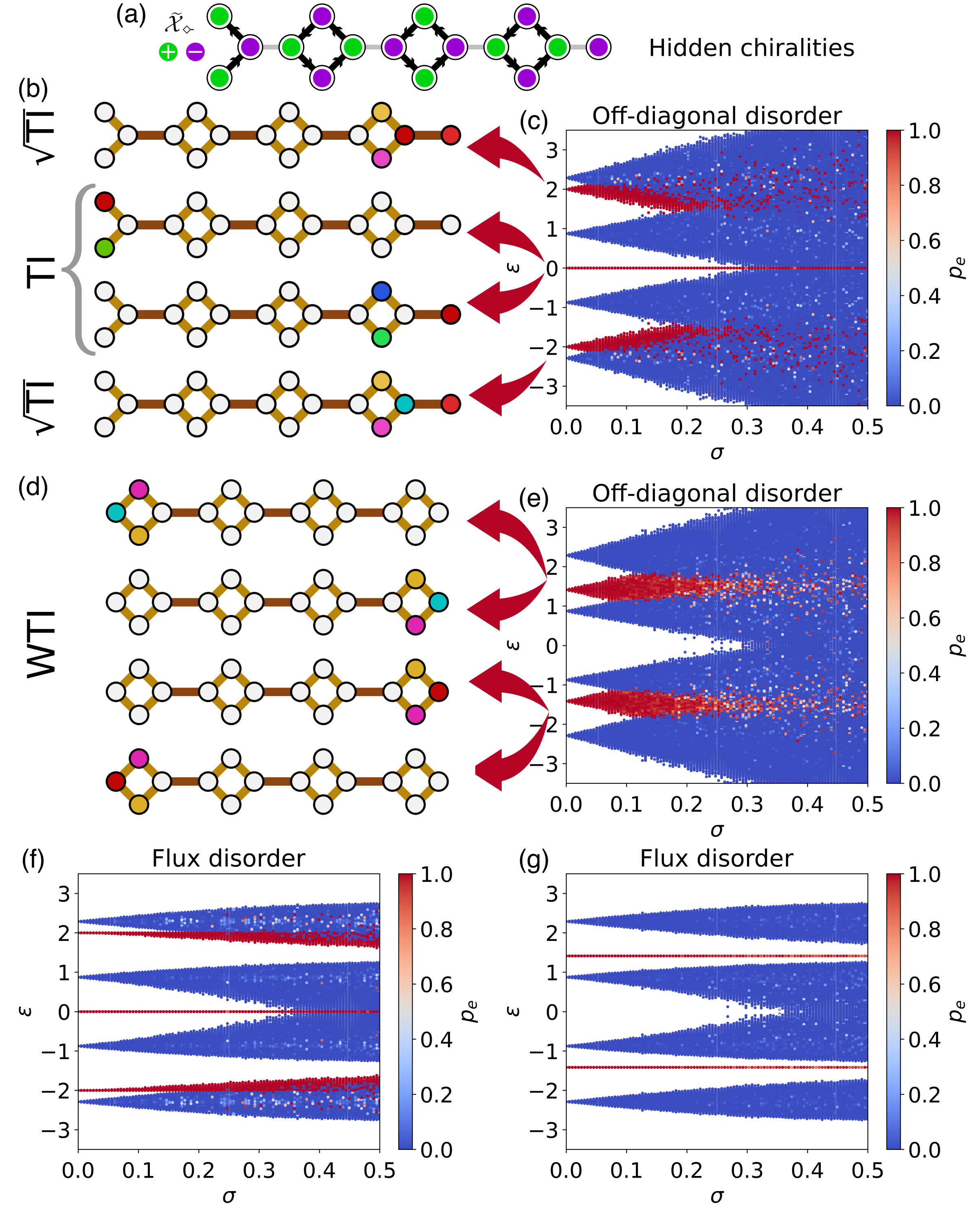}%
}\caption{(a) Chiralities of the hidden symmetry $\tilde\sX_\dnc$ in the DNC. (b) Edge modes of the Open DNC, following the legend in Fig. \ref{fig:CL} (b). The outer (inner) gap states are square-root (strong) states. (c) Spectra of 1000 realizations of the open DNC as a function of the off-diagonal disorder strength $\sigma$, colored as a function of the probability $p_e$ of finding them in the pristine end modes. (d) Edge modes of the closed DNC. All four end modes are weak topological states. (e) Spectra of 1000 realizations of the closed DNC as a function of off-diagonal disorder strength. (f,g) Spectra of 1000 realizations of the open (f) and closed (g) DNC as a function of flux disorder. All simulations have $N=30$, $J=1$, $\Jp=\sqrt{2}$, $\phi=\pi$.
\label{fig:DNC}}
\end{figure}

As can be seen in Fig. \ref{fig:DNC} (a-c), the strong (TI) states are hidden-chiral and protected as long as $\tilde\sX_\dnc$ is preserved. The hidden winding number $\eta$ [see Eq. (\ref{eq:nu})] was calculated numerically for the second gap and it has a nontrivial value of $\eta = -1$ for all values of $\Jp$ with $\phi\ne 0$.

The states in the outer gaps are not topological in general. However, if $\phi = \pi$, they are hidden square-root topological ($\sq$) states. If we square the $\pi$-flux DNC Hamiltonian, we obtain two more copies of itself with renormalized parameters, one with the same boundary conditions (the parent), and another with different ones (the residual model). We describe these models in detail in Appendix \ref{sec:C}.

Just like in the RhC case, defects on one side of the parent model make one of the end modes (the left one) sink into the bands. For this reason, DNC outer gap states are located on the right end only. There are also on-site defects in the upper and lower sites, but they do not disturb the topological end mode, as shown in Fig. \ref{fig:SqBands} (d). If $\phi\ne\pi$, vertical links also appear inside the diamonds, breaking the hidden chiral symmetry and the square-root topology.

The fact that a square-root topological insulator can be its own parent has important consequences in the rich hierarchy (or \textit{family tree}) between topological models that can be established via these square-root (or even $n$-root) relationships \cite{Marques2021}. In particular, it shows that the tree can have closed loops, so its topology could be more complex than previously thought.

\subsection{Closed DNC}
The other choice for the open boundary conditions is to break the $\Jp$ link, keeping all diamonds complete (closed DNC). Then, two states appear in each of the outer gaps, for a total of four end modes [Fig. \ref{fig:DNC} (d,e)]. They are actually weak (WTI) hidden-topological states, that is, they are formed by combining two standard (i.e. strong) hidden-topological modes. This makes them slightly more robust than square-root states, which are instead formed by coupling a strong topological mode with a non-protected state \footnote{In the open DNC, the residual part of the square-root states is actually one of the weak states we describe below, not a trivial state. In any case, they are not symmetry-protected, so the discussion above still holds.}.

We can actually use the same squaring operation as before, now in the opposite direction, to discern the two protected components that make up the weak topological states. We consider one pair of open DNCs, one being the spatial inversion of the other, as shown on the right in Fig. \ref{fig:SqBands} (e). These are the result of squaring the closed DNC. Each of them has two strong states. If we now understand the closed DNC as built from the states in the two open DNCs --as we argued could be done in the RhC discussion-- the strong states only couple to each other and not to the bulk, forming the weak states. We include more details on this decomposition in Appendix \ref{sec:C}.

We show the behavior of the states in both the open and closed DNC in the presence of local off-diagonal and flux disorders. The weak states in the closed DNC are robust against the latter and keep their energy constant for any value of $\phi$, as shown in Fig. \ref{fig:DNC} (g), while the energy of the square-root states in the open DNC varies slightly [Fig. \ref{fig:DNC} (f)]. We explain the weak state protection in detail in Appendix \ref{sec:D}.

\section{Conclusions}
In this paper, we explore a novel type of exotic topological materials, protected by local hidden chiral symmetries. We find their topological invariant and show the added protection LH symmetries can provide, even to states also protected by other symmetries, like those in the rungless $\pi$-flux Creutz ladder.

We also study the topology of the diamond necklace chain in depth, showing it is also a strong hidden topological insulator. Moreover, its outer gap states, less protected, inherit topological features from its central gap states. This marks the first time a square-root or weak topological model has been recognized as its own parent Hamiltonian, as far as we are aware of. Although it is unclear if this phenomenon could also happen without hidden symmetries, the complexity they generate plays a crucial role in this case \footnote{A possible counterargument could be to consider a DNC with different tunneling amplitudes for odd and even rhombi, thus having eight sites per unit cell. The symmetry $\sX_{\dnc}$ would not be hidden then. If such a model was a $\sqrt{\textrm{TI}}$, however, its parent, while also describable as a 8-site DNC, would actually have four sites per primitive unit cell, regaining the hidden symmetry.}.

The end modes we describe can be used to implement fast and robust quantum state transfer \cite{Lang2017,Zurita2023} or entanglement distribution protocols \cite{Hu2020,ZuritaEnt}. The simultaneous presence of several end modes can be used to increase the bandwidth of these protocols. Suitable experimental platforms would be photonic lattices \cite{Mukherjee2018,Kremer2020,He2021}, electronic quantum simulators \cite{Kempkes2023a} or cold atoms \cite{Kang2020,Song2018}.

Additionally, the research shown in this paper has the potential to be the starting point for intriguing new fundamental research. For example, as we will present in a future work \cite{Zurita2024}, hidden symmetries can have the effect of broadening the Wannier functions of the model, which in turn can give rise to a variety of exotic topological states.

\begin{acknowledgments}
We acknowledge enlightening conversations with Marco D\'iaz Maceda, A. M. Marques and R. G. Dias.
C.E.C. was supported by the Spanish MICINN through Grant No. PID2022-139288NB-100. G.P. and J.Z. were supported by Spain's MINECO through Grant No. PID2023-149072NB-I00 and by CSIC Research Platform PTI-001. We also acknowledge support from National Project QTP2021-03-002. J.Z. recognizes the FPU program FPU19/03575.
\end{acknowledgments}


\appendix 

\section{Implementation of the disorder}\label{sec:A}
To implement disorder, we use random numbers $\kappa$ with a normal distribution of zero mean and standard deviation $\sigma$, which we identify as the disorder strength. For \textit{on-site disorder}, the energy of each site $j,\alpha$ is modeled for each realization $i$ as:
\begin{equation}
    \epsilon_{j,\alpha} = \kappa_{j,\a}^{(i)}.
\end{equation}

For \textit{off-diagonal disorder}, we modify each off-diagonal element of the Hamiltonian $\sH_{j,\a;j^\prime,\a^\prime} \equiv \xi \sJ$---with $|\xi|=1$ and $\sJ\in \dR^+$---in the following way:
\begin{equation}
    \sH_{j,\a;j^\prime,\a^\prime} = \xi (\sJ + \kappa_{j,\a;j^\prime,\a^\prime}^{(i)}).
\end{equation}
In order to avoid non-physical sign flips, which would instead be associated with magnetic flux fluctuations, we cutoff the normal distribution on the left at $\kappa=-\sJ$. This will affect results significantly only if $\sigma$ is comparable to the hopping amplitudes or higher, so the effect at low disorder is negligible.

For \textit{vertical-link disorder} in the Creutz ladder, we consider the effect of adding a hopping amplitude of $m=\kappa_j^{(i)}$ between the two sites of each unit cell. This could appear naturally in some implementations like the ones proposed in Refs. \cite{Mukherjee2018,Zurita2021}, where the CL arises when decoupling the spinal sites in a $\pi$-flux rhombus chain. There, magnetic destructive interference is the only reason the two atoms in a rung do not interact ($m=0$), and any fluctuation in the system will change that.

We also calculate the standard deviations of the levels [as shown in Fig. \ref{fig:DisCL} (d)] by labeling each state according to its order in the disordered spectrum. This method has the limitation of mixing levels that are closer to each other than the scale set by the disorder strength, but it is accurate for end modes at disorders less than the energy gap.

Finally, we also consider \textit{flux disorder} in the DNC, where we add a perturbation to each Peierls phase, substituting each diamond link, which had a pristine value of $-Je^{i\phi/4}$, with:
\begin{equation}
    -Je^{i(\phi/4 + \kappa_{j,\a;j^\prime,\a^\prime}^{(i)})}.
\end{equation}

\section{Chern number for square-root TIs}\label{sec:B}
Non-conventional topology in one dimension has sometimes been studied using an extended 2D model, created by adding some parameter $\varphi\in [0,2\pi )$ that interpolates between it and its spatially inverted version (or $\sI$-partner) in a nontrivial way. That parameter is taken as the second component of the quasimomentum. Some topological properties of the 1D model can then be deduced from those of the 2D model. In particular, if the resulting model has a nontrivial Chern number, the end modes of the 1D model can be seen as a slice of its edge modes. If $\varphi$ represented time, the model would describe a topological pumping protocol. This technique can be useful to create new 2D topological models from 1D ones, by deriving the corresponding 2D model in real space \cite{Antao2024}, or to find new pumping schemes.

This technique is often used to characterize quasiperiodic topology (where $\varphi$ is the phasonic degree of freedom) \cite{Kraus2012a,Antao2024} and periodic models without spatial inversion symmetry like the trimer chain \cite{MartinezAlvarez2019a} or the CSSH ladders \cite{Zurita2021}. Some models previously studied only in this way, like the latter, turn out to be square-root topological insulators \cite{Zurita2024}.

\begin{figure}[!tph]
\mbox{%
\includegraphics[width=1\columnwidth]{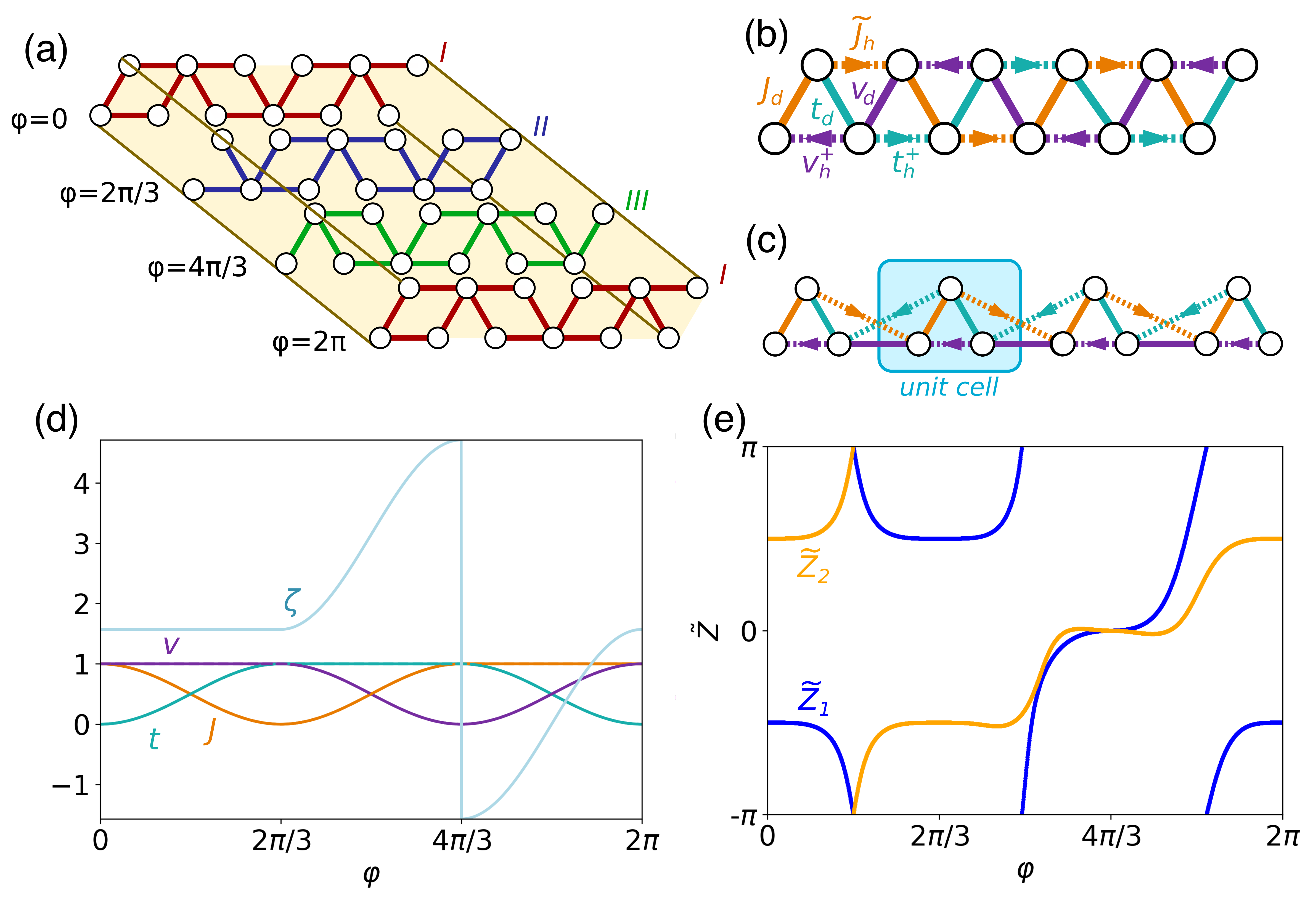}%
}\caption{(a) Effective 2D model, as a function of the periodic parameter $\varphi$. (b) General model used, with the parameters used in the pumping. (c) Demonstration of the 3-periodicity of the model. (d) Pulses used for all parameters. (e) Zak phases of the two gaps.\label{fig:Rh2D}}
\end{figure}

In fact, this technique can be used on any $\sq$, simply by taking the model and its $\sI$-partner and creating a topological pumping cycle with them, i.e. a cycle that translates the Wannier function centers always in the same direction. To illustrate this, we show the corresponding pumping for the $\pi$-flux rhombus chain in Fig. \ref{fig:Rh2D}. We interpolate using $\varphi$ between the rhombus chain (I) at $\vf=0$, its inverted partner (II) at $\vf=2\pi/3$, and an intermediate model (III), needed to complete the cycle, at $\vf=4\pi/3$. We consider a general model, shown in Fig. \ref{fig:Rh2D} (b), where the horizontal hopping amplitudes accumulate all the magnetic phases. The arrows show the direction of the phase shift indicated. We tune the parameters as a function of time as:
\begin{align}
    &t_h^+ = it(\vf),\quad v_h^+ = iv(\vf),\quad\tilde{J_h} = J(\vf) e^{i\zeta(\vf)} \\
    &t_d=t(\vf),\quad  v_d=v(\vf), \quad J_d=J(\vf),
\end{align}
where all parameters are tuned in a sinusoidal way between the three main models, I ($t=0,v=J=1,\zeta=\pi/2$), II ($J=0,t=v=1,\zeta=\pi/2$) and III ($v=0,t=J=1,\zeta=-\pi/2$). This is plotted in Fig. \ref{fig:Rh2D} (d).

The Zak phases associated with both gaps of the model, $\tilde \sZ_{1,2}$, have a winding number of 1 along the domain of $\vf$, as shown in Fig. \ref{fig:Rh2D} (e). This corresponds to a Chern number of 1 for both gaps of the 2D model, which predicts the presence of edge modes in the gaps. The square-root states present in the rhombus chain are cross sections of those states for a fixed $\vf$, and as such, they are not guaranteed to always exist.

\section{Squared DNC Hamiltonians}\label{sec:C}
The squared open DNC Hamiltonian decouples into the subsystems:
\begin{align}
    &\sH_1 = \sH_{\dnc}^{(o)}\left[J\Jp,2J^2\cos\frac{\f}{2}\right] + \sH_{d}   +\nonumber\\
    &- J^2(|1,B\kt\br 1,B| + |1,C\kt\br 1,C|) + J^2e^{i\phi/2}|1,C\kt\br 1,B|,\nonumber\\
    &\sH_2 = \sH_{\dnc}^{(c)}\left[J\Jp,2J^2\cos\frac{\f}{2}\right]+\sH_d -2J^2 |N+1,D\kt\br N+1,D|,\nonumber
\end{align}
where $\sH_{\dnc}^{(o/c)}[J,\Jp]$ is the open/closed DNC in Eq. (\ref{eq:DNC}) with hopping amplitudes $J$ and $\Jp$ ($\phi$ is equal for all models), and
\begin{align}
    &\sH_{d}=\sum_{j} \left[-\Jpsq(|j,B\kt\br j,B| + |j,C\kt\br j,C|)+ \right. \nonumber\\
    & \left.2J^2\cos \frac{\f}{2}|j,B\kt\br j,C|\right] + (\Jpsq+2J^2)\uno
\end{align}
is the Hamiltonian for the defects that appear in every unit cell: a vertical link through each diamond, some on-site energies and a global energy shift $\Delta=(\Jpsq+2J^2)\uno$. For $\phi=\pi$, the vertical link vanishes. In each Hamiltonian $\sH_{1,2}$, the term $\sH_d$ should be understood as affecting only the sites that make up that subspace: as shown in Fig. \ref{fig:SqBands} (d), the open DNC that emerges ($\sH_1$) is formed by the sites at odd positions (white circles), while the closed DNC ($\sH_2$) is formed by those at even positions (grey squares). Due to the way the atoms are labeled, with $A$ being the left atom in each diamond, we consider that the original open DNC starts with atoms $1,B$ and $1,C$ and ends with atom $N+1,A$. This is just a quirk of the labeling scheme: there are still $N$ complete unit cells in the model.

Similarly, the square of the closed DNC decouples into:
\begin{align}
    &\sH_3 = \sH_{\dnc}^{(ro)}\left[J\Jp,2J^2\cos\frac{\f}{2}\right] +\sH_d -\Jpsq |1,A\kt\br 1,A|,\nonumber\\
    &\sH_4 = \sH_{\dnc}^{(o)}\left[J\Jp,2J^2\cos\frac{\f}{2}\right] +\sH_d   - \Jpsq |N,D\kt\br N,D|,\nonumber
\end{align}
where $\sH_{\dnc}^{(ro)}[J,\Jp]$ is the reflected open DNC (starting with a $\Jp$ link and ending in an open diamond). As shown in Fig. \ref{fig:SqBands} (e), $\sH_3$ is formed by the sites in even positions (grey squares), and $\sH_4$ is formed by the rest (white circles).

\section{Weak state protection in the DNC}\label{sec:D}
In the presence of local off-diagonal disorder, the energies of the weak DNC states are not fixed, but they do form symmetric pairs around zero. They are, however, protected by a crystalline symmetry. Finally, they are robust against local flux disorder. Here, we explain these three types of partial protection.

The $\phi=\pi$ closed DNC has the following four edge states:
\begin{align}
    |\sL_{I/\III}\kt = \pm \frac{1}{\sqrt{2}} |1,A\kt + \frac{1+i}{2\sqrt{2}}|1,B\kt + \frac{1-i}{2\sqrt{2}}|1,C\kt \label{eq:Ldnc}\\
    |\sR_{I/\III}\kt = \pm \frac{1}{\sqrt{2}} |N,D\kt + \frac{1-i}{2\sqrt{2}}|N,B\kt + \frac{1+i}{2\sqrt{2}}|N,C\kt, \label{eq:Rdnc}
\end{align}
with energies $\e_I^{(\sL/\sR)} = -\e_\III^{(\sL/\sR)} = \sqrt{2} J$, located in the lower and upper gaps, which we represent with $I$ and $\III$.

In the presence of disorder, the energies of the four states fluctuate, but the symmetries of the model constrain them in two ways. We label the energies $\e_{\tilde n}^{(\mu)}$, with $\ny = I,\III$ labeling the gaps and $\mu=\sL,\sR$ labeling the ends of the model.

The edge states in the two gaps are chiral partners of each other under $\tilde\sX_\dnc$: $\tilde\sX_\dnc |\mu_I\kt = |\mu_\III\kt; \tilde\sX_\dnc |\mu_\III\kt = |\mu_I\kt$. Through the mechanism that usually protects chiral states in 1D TIs, we can conclude that $\e_\III^{(\mu)} =- \e_I^{(\mu)}$, that is, \textbf{the energies for both left end states are symmetric around zero}, and the same goes for the right end states.
The states are not chiral, so their energies are not fixed at zero. This relation will hold for all kinds of off-diagonal disorder, given that $\tilde\sX_\dnc$ is diagonal.

We now consider the \textbf{crystalline (nonlocal) symmetry} $\tilde \sV \equiv \tilde\sU \sI$, where $\sI$ is the 2D inversion symmetry around the center point of the finite model and $\tilde \sU$ is the block-diagonal matrix:
\begin{equation}
    \tilde \sU = \textrm{diag}(1,-\sigma_y,-1,-1,\sigma_y,1,\ldots), \label{eq:sU}
\end{equation}
which repeats every eight rows.

The effect of $\tilde\sV$ on the end modes is: $\tilde\sV |\sL_\ny\kt = (-1)^N|\sR_\ny\kt$; $\tilde\sV |\sR_\ny\kt = |\sL_\ny\kt$. Therefore, we can write:
\begin{align}
    &\sH_\dnc \tilde \sV |\sL_\ny\kt = (-1)^N\sH_\dnc |\sR_\ny\kt = (-1)^N\e_\ny^{(\sR)} |\sR_\ny\kt\nonumber\\
    &= \tilde\sV \sH_\dnc |\sL_\ny\kt = \tilde\sV \e_\ny^{(\sL)} |\sL_\ny\kt = (-1)^N \e_\ny^{(\sL)} |\sR_\ny\kt,\nonumber
\end{align}
i.e. $\e_\ny^{(\sR)} = \e_\ny^{(\sL)}$. The states in each gap have the same energy as long as $\tilde \sV$ is preserved. As usual for crystalline symmetries, however, any amount of local disorder will break it.

Finally, we can explain the \textbf{robustness against local flux disorder}, shown in Fig. \ref{fig:DNC} (g), by noting that changes in $\phi$ do not change the energy of either of the two parent states. Even if the phase of every link fluctuates (which amounts to changing the flux in all diamonds and applying a gauge transformation), the four parent states depicted on the right in Fig. \ref{fig:SqBands} (e) still have an energy of $2J^2$, assuming a long enough chain. Given that they come from squaring the original Hamiltonian, the energy of the weak end modes there has to be $\pm \sqrt{2} J$.

\begin{figure}[!tph]
\mbox{%
\includegraphics[width=\columnwidth]{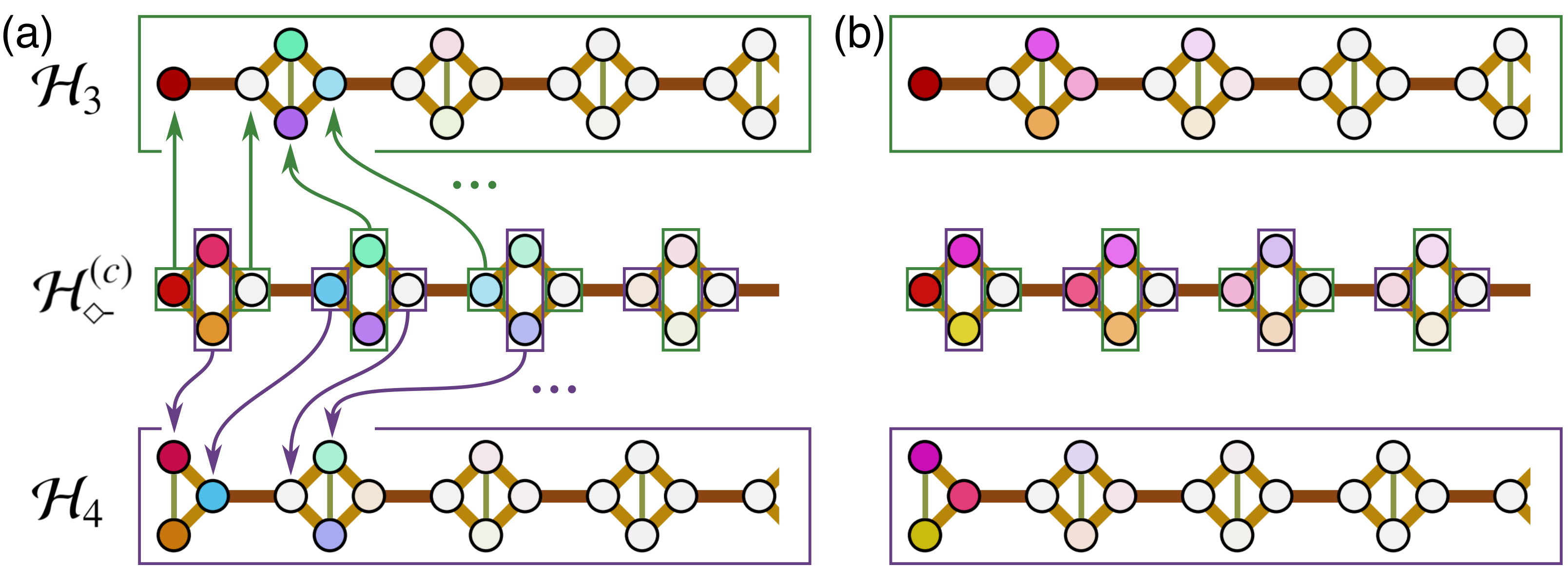}%
}\caption{(a) Form of the weak left end mode at $\ve=-\sqrt{2}J$ in a semi-infinite closed DNC (middle) with flux disorder ($\sigma=0.2$), following the legend in Fig. \ref{fig:CL} (b). The form of the parent states (with an energy of $\ve^2=2J^2$) in the $\sH_{3/4}$ models matches the weak state in the corresponding sites (odd/even $x$-coordinate sites), except for normalization. We show them above/below, with the aforementioned legend. (b) The same states as in (a), in a different disorder realization. This phenomenon is robust against any flux disorder.
\label{fig:weakdphi}}
\end{figure}

The wavefunctions do change with flux disorder and stop being compact. The form of the weak modes is then:
\begin{equation}
    |\sL_{\tilde n}\kt = \sum_j \varkappa_{(j)}^{j-1} \left[\alpha|j,A\kt+\beta^{(j)}_{\ny}|j,B\kt + \gamma^{(j)}_{\ny} |j,C\kt \right],
\end{equation}
where $\beta^{(j)}_{I/\III} = \pm\alpha e^{-i(\phi/4+\delta\phi_{j,1})}/\sqrt{2}$, $\gamma^{(j)}_{I/\III} = \pm\alpha e^{i(\phi/4+\delta\phi_{j,2})}/\sqrt{2}$, $\varkappa_{(j+1)}=\mp J(e^{-i(\phi/2+\delta\phi_{j,1}+\delta\phi_{j,3})} + e^{i(\phi/2 + \delta\phi_{j,2}+\delta\phi_{j,4})})/(\sqrt{2}\Jp)$, and $\alpha \in \dR^+$ is found by normalization. In these expressions, $\delta\phi_{j,i}$ with $i=1,2,3,4$ refers to the disorders in the upper left, lower left, upper right and lower right sides on the $j$-th diamond, respectively. As we show in Fig. \ref{fig:weakdphi}, the weak states' wavefunction in a closed DNC with flux disorder still matches their parent states.



%

\end{document}